\documentclass{webofc}
\usepackage[varg]{txfonts}   
\usepackage{graphics}
\begin{document}
\title{A New Heavy Flavor Program for the Future Electron-Ion Collider}
%
%

\author{\firstname{Xuan} \lastname{Li}\inst{1}\fnsep\thanks{\email{xuanli@lanl.gov} },
        \firstname{Ivan} \lastname{Vitev}\inst{1}, \firstname{Melynda} \lastname{Brooks}\inst{1}, \firstname{Lukasz} \lastname{Cincio}\inst{1}, \firstname{J. Matthew} \lastname{Durham}\inst{1}, \firstname{Michael} \lastname{Graesser}\inst{1}, \firstname{Ming X.} \lastname{Liu}\inst{1}, \firstname{Astrid} \lastname{Morreale}\inst{1}, \firstname{Duff} \lastname{Neill}\inst{1}, \firstname{Cesar} \lastname{da Silva}\inst{1}, \firstname{Walter E.} \lastname{Sondheim}\inst{1}, \firstname{Boram} \lastname{Yoon}\inst{1}.
}

\institute{Los Alamos National Laboratory
          }

\abstract{%
 The proposed high-energy and  high-luminosity Electron-Ion Collider (EIC) will provide one of the cleanest environments to precisely determine the nuclear parton distribution functions (nPDFs) in a wide $x$-$Q^{2}$ range. Heavy flavor production at the EIC provides access to nPDFs in the poorly constrained high Bjorken-$x$ region, allows us  to study the quark and gluon fragmentation processes, and constrains parton energy loss in cold nuclear matter. Scientists at  the Los Alamos National Laboratory are developing a new physics program to study heavy flavor production, flavor tagged jets, and heavy flavor hadron-jet correlations in the nucleon/nucleus going direction at the future EIC. The proposed measurements will provide a unique way to explore the flavor dependent fragmentation functions and energy loss in a heavy nucleus. They will constrain the initial-state effects that are critical for the interpretation of previous and ongoing heavy ion measurements at the Relativistic Heavy Ion Collider  and the Large Hadron Collider. We show an initial conceptual design of the proposed Forward Silicon Tracking (FST) detector at the EIC, which is essential to carry out the heavy flavor measurements. We further present initial feasibility studies/simulations of heavy flavor hadron reconstruction using the proposed FST. 
}
\maketitle
\section{Introduction}
\label{intro}
Heavy flavor production is an ideal probe to study the full evolution of the nuclear medium created in heavy ion collisions at the Relativistic Heavy Ion Collider (RHIC) and the Large Hadron Collider (LHC). Heavy quarks are produced early on in the scattering process due to their high masses ($m_{c/b} \gg \Lambda_{QCD}$). The precise mechanisms by which heavy flavor interacts with the nuclear medium, however,  are still not well understood in heavy ion collisions. The production process includes both  cold nuclear matter effects such as  modification of the nuclear Parton Distribution Functions (nPDF), and hot nuclear matter effects such as parton energy loss  within the Quark Gluon Plasma (QGP). The future Electron Ion Collider (EIC), recommended as top priority for new construction by the 2015  Long Range Plan for nuclear science  and strongly endorsed by the 2018 National Academy of Sciences committee,  will utilize  high-energy high-luminosity electron-nucleon (electron-nucleus) collisions to study several fundamental questions in the nuclear physics field. The deep inelastic scattering (DIS) processes of electrons on light/heavy nuclei at the future EIC will provide a clean environment to directly explore the initial-state cold nuclear matter effects and the final-state fragmentation/hadronization in the nuclear environment. These studies will improve our understanding of the current and future RHIC and LHC measurements and shed light on the non-perturbative aspects of quantum chromodynamics (QCD).

\section{A new heavy flavor and jet program for the future EIC}
\label{sec-1}
The future EIC will provide electron+nucleus collisions with multiple nuclear species ranging from $^{3}He$ to $^{208}Pb$ at a series of center-of-mass energies from 20 GeV to 141 GeV \cite{ref_white_paper}. Unlike heavy ion experiments at RHIC and LHC, these asymmetric collisions will provide a clean environment to precisely study the initial-state nPDFs  in the large Bjorken-x region \cite{npdf1, npdf2} and the fragmentation/hadronization processes for heavy flavor production. Heavy flavor measurements in polarized electron+nucleon/nuclei collisions at the future EIC will provide opportunities to study the nucleon/nucleus spin structure such as the gluon Sivers effect. 

\subsection{Exploration of the flavor-dependent parton energy energy loss  and in medium hadronization at the EIC}
\label{sec-1-1}
\begin{figure}[h]
\centering
\includegraphics[width=6.5cm,clip]{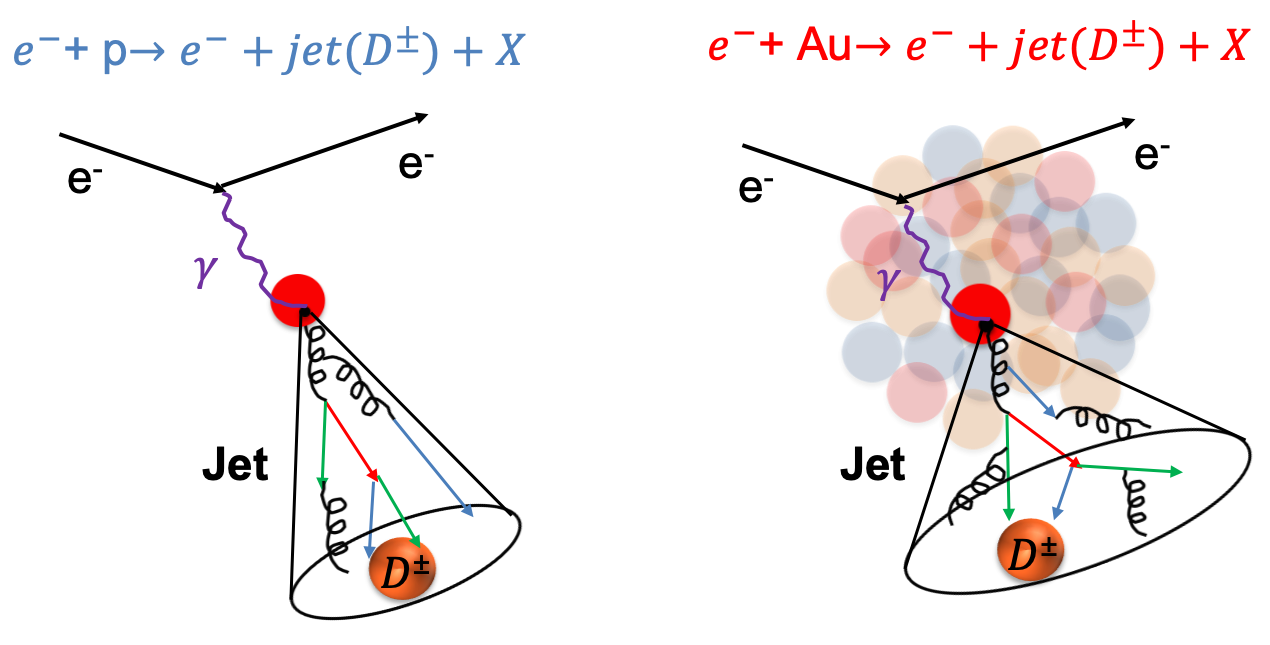}
\caption{Heavy flavor hadron and jet production in electron+proton (left) and electron+nuclei (right) collisions. The heavy flavor fragmentation and hadronization process in electron+nuclei collisions is modified when compared to the process in electron+proton collisions.}
\label{fig-1} 
\end{figure}

Quarks and gluons lose energy when traversing a nuclear medium and their energy loss is mass/flavor dependent ($\Delta E_{g} > \Delta E_{u,d,s} > \Delta E_{c} > \Delta E_{b}$) \cite{RefA}. The quark/gluon fragmentation and hadronization processes can be modified in the nuclear medium as shown in figure~\ref{fig-1}. The cross sections of final-state hadrons and jets to be measured in electron+proton and electron+nuclei collisions at the future EIC are proportional to the initial quark/gluon PDFs, the quark/gluon hard scattering part which can be calculated in perturbative QCD, and the quark/gluon fragmentation and hadronization processes. Comparison of different final-state cross sections can help separate the information about the initial-state nuclear PDFs and the final-state fragmentation and hadronization processes. Comparison of cross sections measured in electron+proton and electron+nucleus collisions for the same final hadron states can precisely determine the nuclear medium effects such as the parton energy loss contributions.

\begin{figure}[h]
\centering
\includegraphics[width=5.2cm,clip]{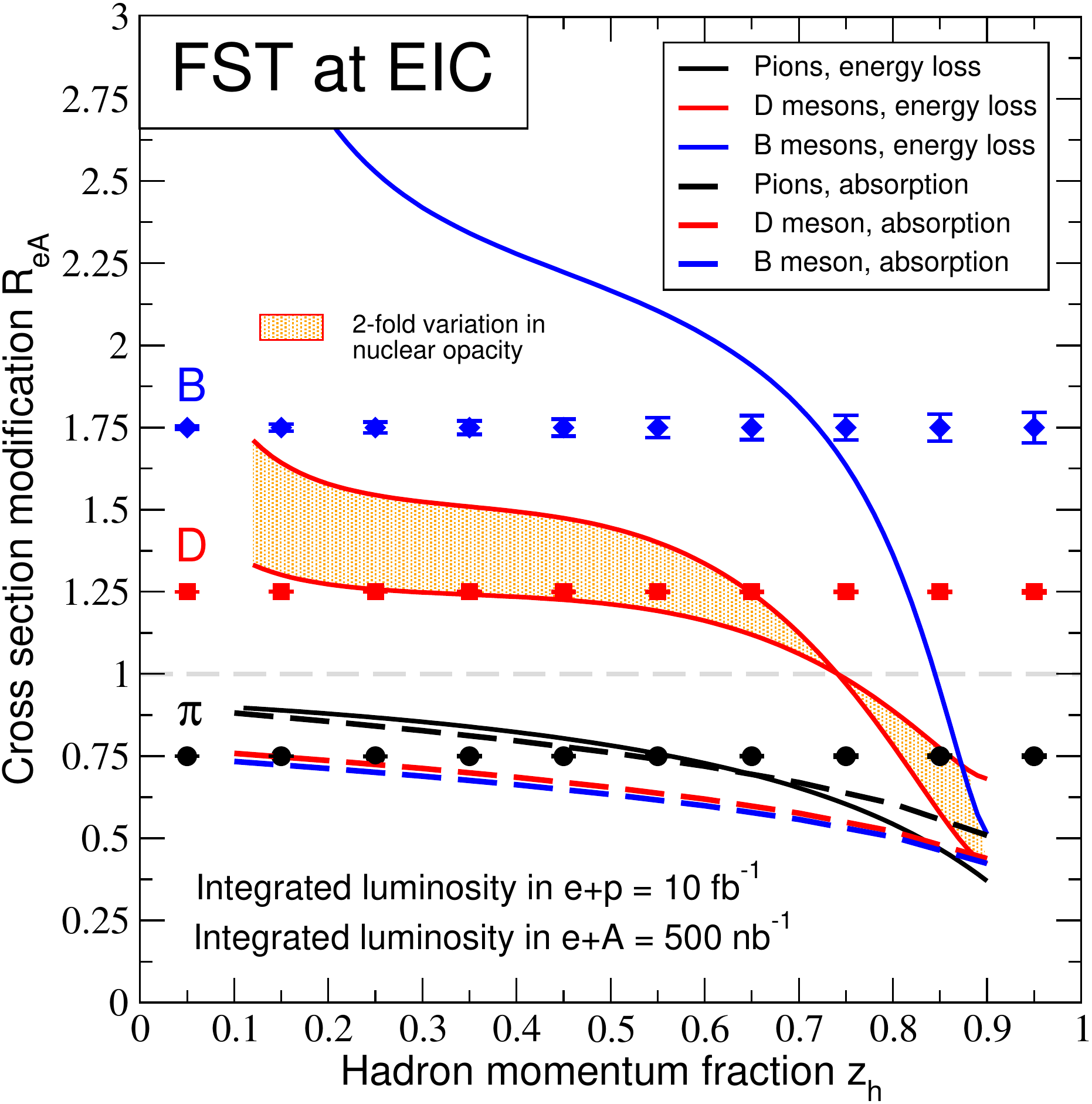}
\caption{Projected statistical uncertainties of the nuclear modification factor $R_{eA}$ for pions, $D$-mesons and $B$-mesons in electron+gold collisions at $\sqrt{s}$ = 69 GeV. The projected $R_{eA}$ mean value is based on assumptions. Theory calculations include  parton energy loss (solid lines) and absorption within nuclei (dashed lines) \cite{RefB}. The integrated luminosity in electron+proton (electron+gold) collisions is 10 $fb^{-1}$ (500 $nb^{-1}$).}
\label{fig-2} 
\end{figure}

There are competing theoretical models of nuclear modification in DIS reactions within nuclei. One is the absorption model, which describes  hadronization that happens inside nuclear matter. The other is quark/gluon energy loss, which is associated with hadronization outside the nuclear medium. Differentiation between these two models is challenging with only light flavor hadron measurements. Figure~\ref{fig-2} shows the predicted nuclear modification factor $R_{eA}$ for heavy flavor hadron measurements versus the hadron momentum fraction relative to their parent parton in $e + Au$ collisions at $\sqrt{s} = 69$ GeV. We show nuclear absorption estimates by dashed lines and  parton energy loss calculations by solid lines. The statistical uncertainties are evaluated at the particle generation level with estimated sampling and reconstruction efficiencies in the forward pseudorapidity region of $1.0<\eta<4.5$, which is the proposed Forward Silicon Tracking coverage. The integrated luminosity in electron+proton (electron+gold) collisions is 10 $fb^{-1}$ (500 $nb^{-1}$). Clear separation between these two models can only be achieved through heavy flavor hadron measurements, as they have very different fragmentation functions and formation time when compared to light flavor hadrons \cite{RefB}. The future EIC with the proposed FST detector can provide sufficient statistical precision for these heavy flavor measurements with enhanced sensitivity to nuclear transport properties.


\subsection{Initial design of the proposed Forward Silicon Tracking detector and its performance}
\label{sec-1-2}
\begin{figure}[h]
\centering
\includegraphics[width=10.5cm,clip]{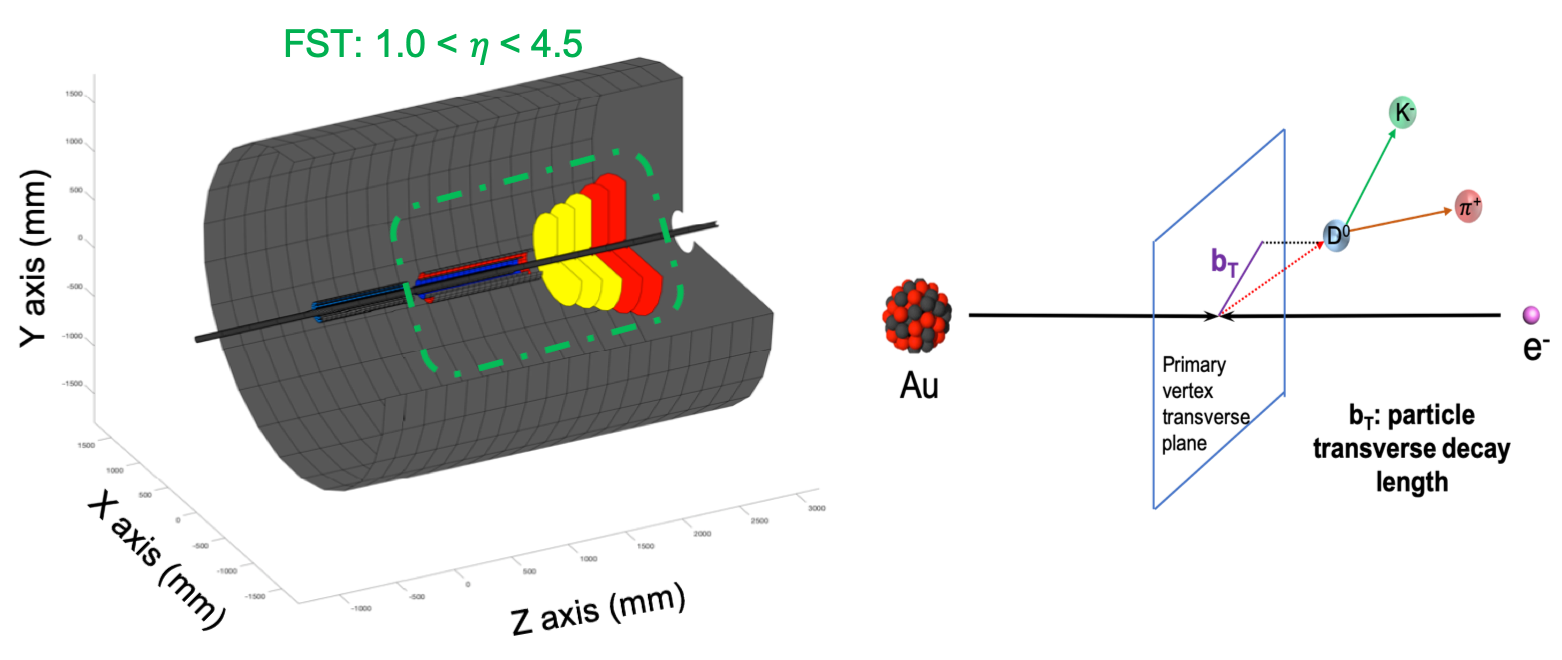}
\caption{Initial design of the proposed Forward Silicon Tracker, which covers the pseudorapidity of $1.5<\eta<4.5$ in the nucleon/nuclei going direction at the future EIC, is highlighted inside the green box on the left. This detector can precisely determine the transverse decay length of particles from the primary vertex ($b_{T}$) as shown on the right.}
\label{fig-3} 
\end{figure}

\begin{figure}[h]
\centering
\includegraphics[width=10.5cm,clip]{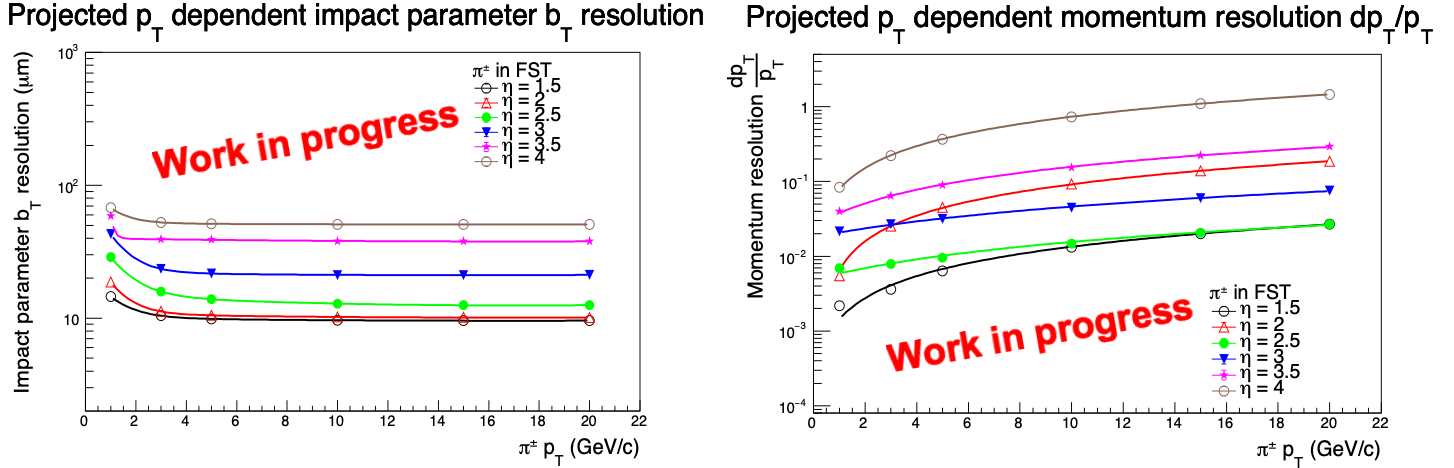}
\caption{Single track performance for the proposed Forward Silicon Tracker in different pseudorapidity regions from $\eta = 1.5$ to $\eta = 4.0$. A uniform 3T magnetic filed is applied for the track performance evaluation. The resolution of the transverse decay length $b_{T}$ versus the track $p_{T}$ is shown on the left and the track $p_{T}$ dependent momentum resolution $\Delta p_{T}/p_{T}$ is shown on the right.}
\label{fig-4} 
\end{figure}

The initial design of the proposed Forward Silicon Tracking (FST) detector in fast simulation is shown in figure~\ref{fig-3}. The proposed FST consists of 2 barrel layers and 5 forward planes to cover the pseudorapidity range of $1.0<\eta<4.5$ in the nucleon/nucleus going direction at the future EIC. This detector can precisely determine the transverse decay length of particles (see the schematics in figure~\ref{fig-3}), which is critical for heavy flavor decay product reconstruction and tagging. Single track performance based on a 3 layer mid-rapidity Monolithic Active Pixel Sensor (MAPS) \cite{maps} central vertex detector and the proposed FST with a hybrid design of MAPS plus other silicon technology is shown in figure~\ref{fig-4}. A uniform 3T magnetic filed is applied in the fast simulation and the magnetic field provided by the EIC interaction point design will be included in the full simulation later. With the help of the FST, better than 70 $\mu m$ spatial resolution can be obtained for forward tracks. Additionally, charged tracks with low transverse momentum can get better than 2$\%$ momentum resolutions.

\subsection{Heavy flavor hadron reconstruction with the proposed Forward Silicon Tracking  detector}
\label{sec-1-3}

To demonstrate the proposed FST capability, the track performance discussed in Sect.~\ref{sec-1-2} is embedded into the PYTHIA8 \cite{py8} simulation for hadron reconstruction studies. One $e+p$ collision is embedded with on average of 0.02 $p+p$ collisions, which account for the beam interaction background. As the EIC background evaluation is underway, we will include other backgrounds such as the synchrotron radiation in future studies. Charm mesons including $D^{0}$, $D^{\pm}$ and $D_{s}^{\pm}$ can decay into $K^{\pm}$, $\pi^{\pm}$ or lighter $D$-mesons with a relatively large branching ratio. The reconstruction algorithm in simulation is developed to cluster charged tracks by matching their transverse decay length ($b_{T}$) ($\Delta b_{T} < 50 \mu m$) measured by the FST (see the geometry in figure~\ref{fig-3}). To enhance the probability of finding charm mesons, the charged track clusters are required to contain at least one $K^{\pm}$ track with 100$\%$ PID. 

\begin{figure}[h]
\centering
\includegraphics[width=9.5cm, clip]{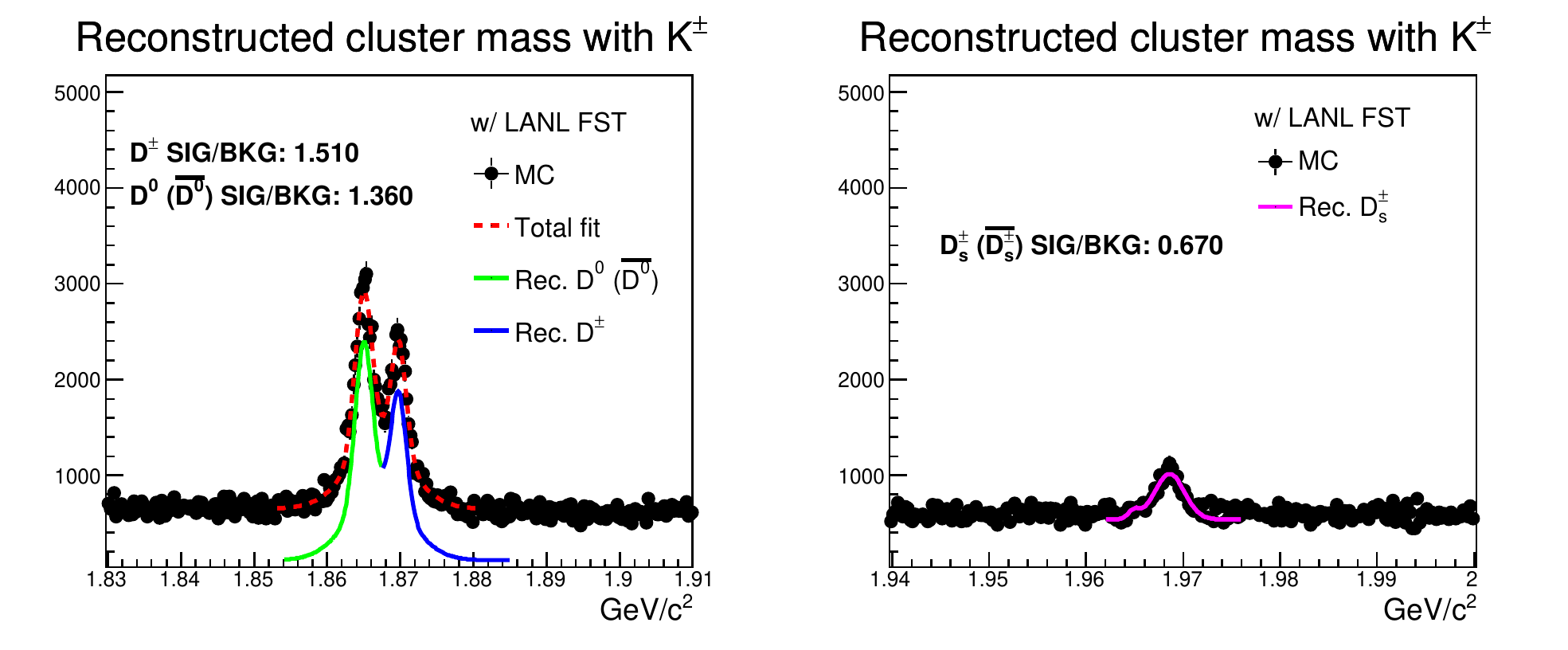}
\includegraphics[width=9.5cm, clip]{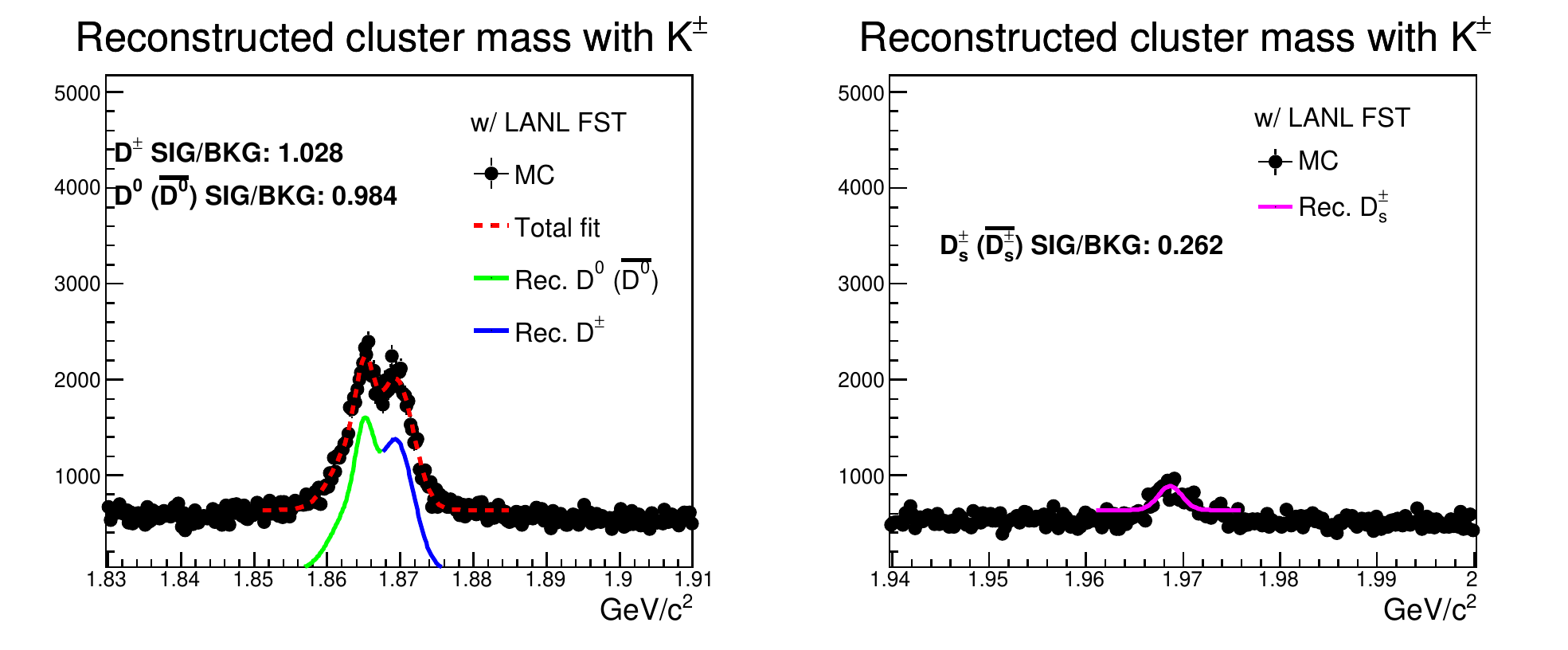}
\caption{Invariant mass of charged track clusters with at least one $K^{\pm}$ track in 18 GeV electron on 100 GeV proton collisions. No charge separation is applied. The integrated luminosity is 10 $fb^{-1}$. Clear $D^{0}$ ($\bar{D^{0}}$), $D^{\pm}$ and $D_{s}^{\pm}$ signals are found with the help of the proposed FST. The performance in the top panel is based on the proposed FST has 30 $\mu m$ pixel pitch, 500 kHZ trigger rate and $0.4 \%$ $X_{0}$ radiation length per detector layer. The distributions in the bottom panel are based on the proposed FST has 200 $\mu m$ pixel pitch, 500 kHZ trigger rate and $0.4 \%$ $X_{0}$ radiation length per detector layer.}
\label{fig-5} 
\end{figure}

Figure~\ref{fig-5} shows the invariant mass distributions of charged track clusters which includes at least one $K^{\pm}$ track in 18 GeV electron and 100 GeV proton collisions in different mass regions. Two different FST pixel pitch values are used to compare the reconstructed $D$-meson mass distributions. With 10 $fb^{-1}$ integrated luminosity, clear signals of $D^{0}$ ($\bar{D^{0}}$), $D^{\pm}$ and $D_{s}^{\pm}$ are found with the help of the proposed FST, which has 30 $\mu m$ pixel pitch, 500 kHZ trigger rate and $0.4 \%$ $X_{0}$ radiation length per detector layer. The signal to background ratio for $D$-meson reconstruction is determined from a fit which includes the signal and combinatorial background contributions. When the pixel pitch increases to 200 $\mu m$, the mass resolution for the reconstructed $D$-mesons gets broader and the signal to background ratio is reduced.

To study the physics results, such as the $D$-meson reconstruction, dependence  on the detector performance, we scan the 3D parameter space which includes the pixel pitch, the trigger integration time, and the average material budgets per layer. Table~\ref{tab-1} summarizes the parameter values in the simulation studies.

\begin{table} [h]
\centering
\caption{The proposed FST parameter table}
\label{tab-1}       
\begin{tabular}{lllll}
\hline
Parameter name &  &  values &  &  \\ \hline
Pixel pitch & 30 $\mu m$ & 50 $\mu m$ & 100 $\mu m$ & 200 $\mu m$  \\ \hline
Material budgets per layer & 0.3$\%$ $X_{0}$ & 0.4$\%$ $X_{0}$ & 0.5$\%$ $X_{0}$  & 1.0$\%$ $X_{0}$  \\ \hline
Trigger integration time & 2 $\mu s$ & 4 $\mu s$ & 8 $\mu s$ &   \\ \hline
\end{tabular}
\end{table}

\begin{figure}[h]
\centering
\includegraphics[width=4.5cm,clip]{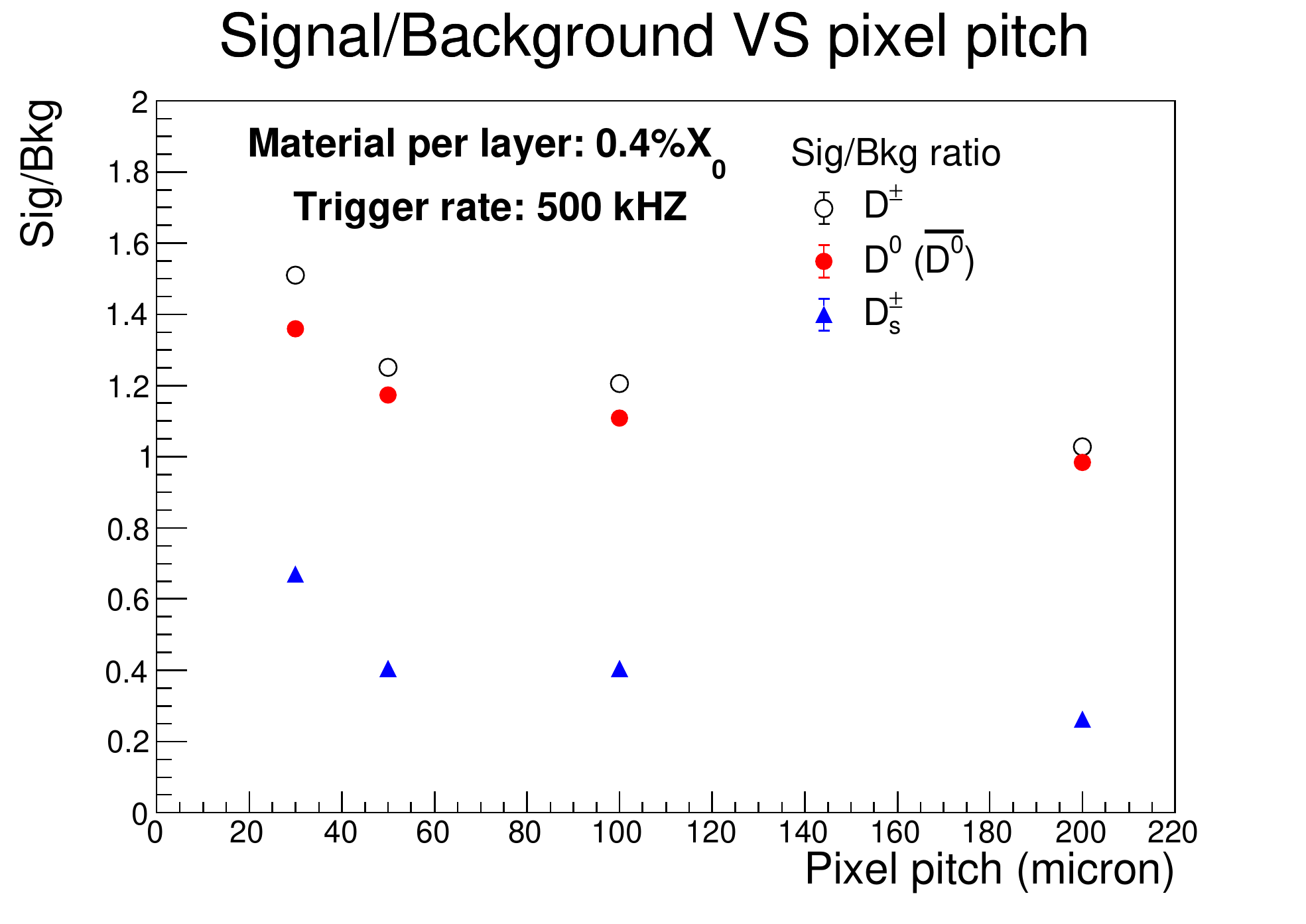}
\includegraphics[width=4.5cm,clip]{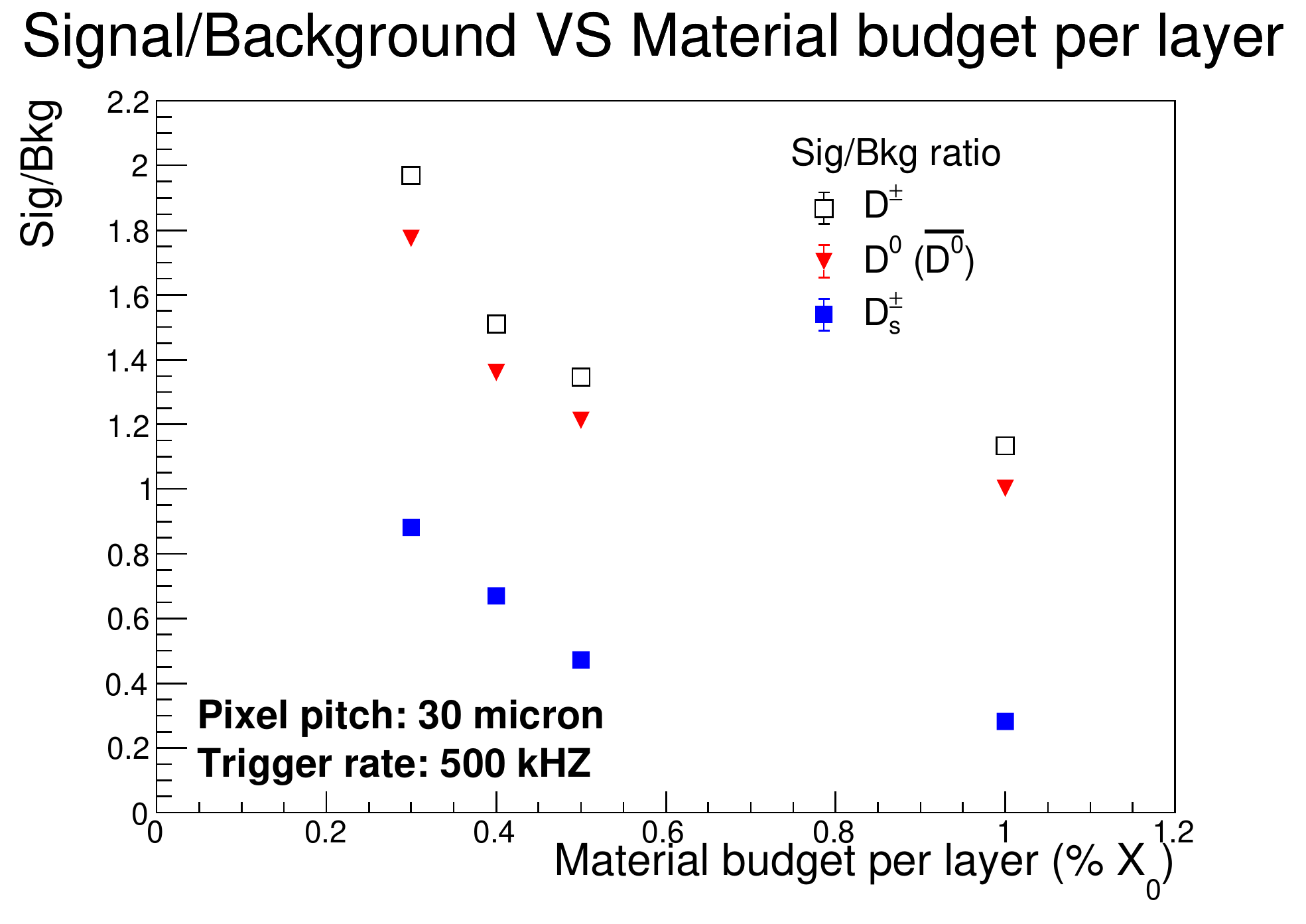}
\includegraphics[width=4.5cm,clip]{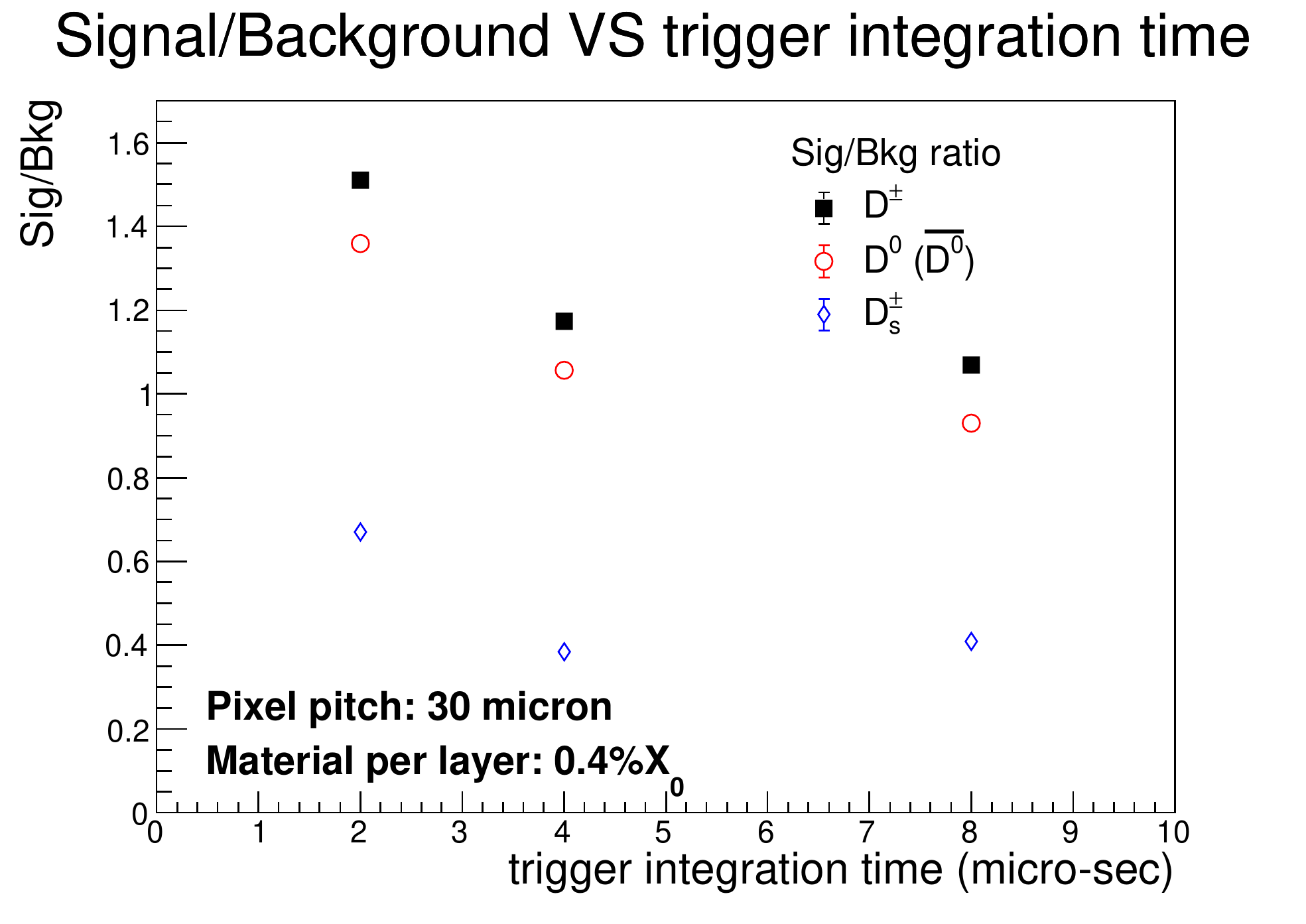}
\caption{Signal to background ratio for reconstructed $D^{0}$ ($\bar{D^{0}}$), $D^{\pm}$ and $D_{s}^{\pm}$ versus pixel pitch (left), materials budgets per detector layer (middle) and the trigger integration time (right) for different detector performance combinations. These reconstruction values are achieved with 10 $fb^{-1}$ integrated luminosity.}
\label{fig-6} 
\end{figure}

By varying the pixel pitch, the material budget per layer and the trigger integration time, the extracted signal over background ratios for reconstructed $D^{0}$ ($\bar{D^{0}}$), $D^{\pm}$ and $D_{s}^{\pm}$ in 18 GeV electron on 100 GeV proton collisions are shown in figure~\ref{fig-6}. The reconstruction statistical uncertainties, which are determined with 10 $fb^{-1}$ integrated luminosity and include 95$\%$ tracking efficiency, are better than 0.1$\%$.

The $D$-meson reconstruction signal to background ratio decreases as pixel pitch, material budgets per detector layer and trigger integration time increases. It has a strong dependence on the material budgets per detector layer with fixed pixel pitch at 30 $\mu m$ and fixed trigger rate at 500 kHZ. Further studies include a wider detector parameter phase space scan, higher mass hadron (e.g. $B$-mesons) and jet reconstruction, more realistic interaction point design and backgrounds in full simulation are underway.

\section{Summary and Outlook}
\label{sec-3}
The future EIC will provide unique opportunities to utilize heavy flavor and jet probes to precisely determine the nuclear PDFs, explore the gluon Sivers function and study parton energy loss mechanism in nuclear medium within a wide Bjorken-x and $Q^{2}$ kinematic region. A new heavy flavor and jet program has started at Los Alamos National Laboratory (LANL) to develop new observables and initialize a conceptual detector design for a proposed Forward Silicon Tracking detector  for the future EIC. Heavy flavor production in $e+A$ collisions at the future EIC will provide strong discriminating power between competing theoretical models for the parton transport coefficients in nuclear medium. A fast simulation version of the FST design is completed and its tracking performance is consistent with or better than the EIC detector R$\&$D handbook requirements \cite{eic_handbook}. Initial studies in the fast simulation demonstrate the capability for $D$-meson reconstruction by the proposed FST. A silicon detector R$\&$D lab is being set up to verify the performance of the proposed silicon techniques, such as MAPS. Ongoing theoretical and experimental studies at LANL, which include new heavy flavor and jet observables determination, physics projections in full simulation, detector design optimization and system integration, will bring about key advances in theory and experimentation and provide important contributions to the EIC physics and detector developments.

%
%
\bibliographystyle{iopart-num}

\end{document}